\documentclass[11pt]{article}

\usepackage{graphicx}
\usepackage{psfrag}
\usepackage{amsmath}
\usepackage{amssymb}
\usepackage{dcolumn}

\newcolumntype{d}[0]{D{.}{.}{-1}}

\newcommand{\newhline}[0]{\\[1mm] \hline}
\newcommand{\slashed}[1]{\ensuremath{{#1}{\!}{\!}{\!}{\!}{\:}/}}
\newcommand{\Slashed}[1]{\ensuremath{{#1}{\!}{\!}{\!}{\!}{\!}{\:}/}}

\newcommand{\eg}[0]{\textit{e.g.}}

\newcommand{\citereference}[1]{Ref.~\cite{#1}}
\newcommand{\citetworeferences}[2]{Refs.~\cite{#1}~and~\cite{#2}}

\newcommand{\citereferencedequation}[2]{equation~({#2}) in \citereference{#1}}
\newcommand{\citethreereferencedequations}[4]{equations~({#2}), ({#3}) and ({#4}) in \citereference{#1}}
\newcommand{\citefigure}[1]{Fig.~\ref{#1}}
\newcommand{\citetable}[1]{Tab.~\ref{#1}}

\newcommand{\texttanbeta}[0]{$\tan\beta$}
\newcommand{\tautothreemu}[0]{$\tau \to \mu \mu \bar{\mu}$}
\newcommand{\tautomugamma}[0]{$\tau \to \mu \gamma$}
\newcommand{\taulepton}[0]{$\tau$ lepton}
\newcommand{\tauleptons}[0]{$\tau$ leptons}
\newcommand{\cosTheta}[0]{$\cos\Theta$}
\newcommand{\differentiald}[0]{\ensuremath{\text{d}}}

\newcommand{\gammabar}[0]{\ensuremath{{\gamma}^{5}}}

\newcommand{\superpartner}[1]{\ensuremath{{\tilde{{#1}}}}}
\newcommand{\mydefinedby}[0]{\ensuremath{{\:}{\equiv}{\:}}}

\newcommand{\mysin}[1]{\ensuremath{{\sin}{\!}{\!}{\:}{\!}{\:}\left( {#1} \right)}}
\newcommand{\mycos}[1]{\ensuremath{{\cos}{\!}{\!}{\:}{\!}{\:}\left( {#1} \right)}}

\begin{document}

\baselineskip=18pt

\begin{titlepage}

\begin{flushright}
{\footnotesize{PITHA 08/04}\\
\today}
\end{flushright}

\vspace{0.6cm}

\begin{center}
{\Large \bf The lepton--flavour violating decay \begin{boldmath} ${\tau}
    {\to} {\mu} {\mu} {\bar{{\mu}}}$ \end{boldmath} at the LHC}\\[1cm]
{\large M.~Giffels$^1$, J.~Kallarackal$^{2,3}$, M.~Kr{\"{a}}mer$^2$, B.~O'Leary$^2$, 
and A.~Stahl$^{1}$}\\[1cm]
{\it $^1$ III. Physikalisches Institut, RWTH Aachen, 52056 Aachen, Germany\\[2mm]
     $^2$ Institut f\"ur Theoretische Physik, RWTH Aachen, 52074 Aachen, Germany\\[2mm]
     $^3$ Institut f\"ur Physik, Humboldt--Universit\"at zu Berlin, 12489 Berlin, Germany}
\end{center}

\vspace{1.8cm}

\begin{abstract}
\baselineskip=18pt
  Lepton--flavour violating ${\tau}$--decays are predicted in many
  extensions of the Standard Model at a rate observable at future
  collider experiments.  In this article we focus on the decay ${\tau}
  {\to} {\mu} {\mu} {\bar{{\mu}}}$, which is a promising channel to
  observe lepton--flavour violation at the Large Hadron Collider LHC.
  We present analytic expressions for the differential decay width
  derived from a model--independent effective Lagrangian with general
  four--fermion operators, and estimate the
  experimental acceptance for detecting the decay ${\tau} {\to} {\mu}
  {\mu} {\bar{{\mu}}}$ at the LHC.  Specific emphasis is given to
  decay angular distributions and how they can be used to discriminate
  new physics models.  We provide specific predictions for various
  extensions of the Standard Model, including supersymmetric, little
  Higgs and technicolour models.
\end{abstract}

\end{titlepage}


\section{Introduction}\label{sec:intro}

In the Standard Model (SM) with massless neutrinos, lepton flavour is
conserved.  However, the current neutrino oscillation data indicate
non--degenerate neutrinos with large mixing angles~\cite{PDG}, and
this, in turn, implies lepton--flavour violation (LFV) within the SM
extended to include massive neutrinos.  This flavour violation is
large in the neutrino sector (where the
Pontecorvo--Maki--Nakagawa--Sakata (PMNS) mixing matrix has large
off--diagonal entries, unlike the Cabbibo--Kobayashi--Maskawa (CKM)
matrix of the quark sector), but very small in the charged--lepton
sector, with branching ratios for LFV decays suppressed by factors of
${\delta}m_{{\nu}}^{2} / m_{W}^{2}$~\cite{Lee:1977tib}, and thus well
below current and future experimental limits.  Hence, any experimental
signal of charged--lepton--flavour violation would be a clear
indication of physics beyond the SM. While current bounds from
non--collider experiments strongly constrain $\mu \to e$ transitions,
the limits on $\tau \to \mu$ and $\tau \to e$ conversion are much less
stringent~\cite{PDG}.  Moreover, as we shall discuss in detail below,
many extensions of the SM predict LFV in $\tau$ decays at a rate
accessible at future $e^{+}e^{-}$ and hadron colliders.

In this article we focus on the decay \tautothreemu\, which should
provide a clean signature to observe LFV at the Large Hadron Collider
LHC~\cite{Unel:2005fj}.\footnote{LFV can also be probed at a future
  $e^+e^-$ linear collider~\cite{Deppisch:2003wt} and at muon or
  neutrino factories~\cite{Sher:2003vi}.}  In general, the decay
${\tau} {\to} e e {\bar{e}}$ would have similar characteristics,
though in many models (including several of the models we consider),
the couplings to electrons are suppressed relative to the couplings to
muons. Moreover, muons provide a far cleaner signal than electrons in
a hadron collider environment~\cite{Unel:2005fj}.  At the LHC, $\tau$
leptons are produced predominantly from decays of $B$ and $D$ mesons
and $W$ and $Z$ bosons. In the low--luminosity phase, corresponding to
an integrated luminosity of $10$~fb$^{-1}$ per year, one expects
approximately $2\times 10^{12}$ and $2\times 10^8$ $\tau$ leptons
produced per year from heavy meson and weak boson decays,
respectively.  The $\tau$ leptons from heavy meson decays result in a
much softer muon transverse momentum spectrum and are more difficult
to trigger and analyze.  Therefore, in our Monte Carlo studies we only
include $\tau$ leptons from $W$ and $Z$ boson decays. With standard
acceptance cuts ($|\eta_{\mu}| < 2.5$ and $p_{{\rm T},\mu} > 3$~GeV),
and requiring either two muons with $p_{\rm T} > 7$~GeV or a single
muon with $p_{\rm T} > 19$~GeV for trigger purposes, we find
acceptances for various BSM models of approximately $25$--$30\%$.
Thus, even restricting ourselves to $\tau$ leptons from weak boson
decays only, and assuming a branching ratio close to the current upper
limit ${\rm BR}(\mbox{\tautothreemu}) \le 1.9 \times
10^{-7}$~\cite{PDG}, we can expect approximately $2 \times 10^{8}
\times 25\% \times 1.9 \times 10^{-7} \simeq 10$ \tautothreemu\ events
within the acceptance range of a typical LHC general purpose detector
after one year of low--luminosity running.  With $30$~fb$^{-1}$ of
data, it should be possible to probe branching ratios down to a level
of ${\rm BR}(\mbox{\tautothreemu}) \approx
10^{-8}$~\cite{Unel:2005fj,CMS_note} at the LHC.  [Recent results from
$B$ factories also begin to test branching ratios down to the level of
$5\times 10^{-8}$~\cite{Abe:2007ev}; a prospective super~B
facility~\cite{:2007zzg} could probe branching ratios of $ {\cal
  O}(10^{-10})$.]

Note that there is an even tighter experimental limit on the radiative
LFV decay \tautomugamma, with an upper bound on the branching ratio of
${\rm BR}(\mbox{\tautomugamma}) \le 6.8 {\times} 10^{-8}$~\cite{PDG}.
The radiative decay itself is difficult to detect at the
LHC~\cite{Unel:2005fj}, and any model that allows the decay
\tautothreemu\ to proceed through an effective ${\tau} {\mu} {\gamma}$
vertex only is already constrained by this bound, and will be further
suppressed by a factor of ${\alpha}$ from the ${\bar{{\mu}}} {\mu}
{\gamma}$ vertex. However, models where the LFV $\tau$ decay can be
mediated by the exchange of new heavy particles escape the tight bound
on radiative transitions. With this in mind, we will only investigate
models of this type, though we will briefly comment on other models
afterwards.

There are many observables associated with the decay \tautothreemu.
Since we are focusing on models where the virtual particles have large
masses $M\gg m_\tau$, the propagators are replaced by $ -i / M^{2}$,
and the obvious difference between models is the chiral structure of
the effective vertices.  We choose the angle ${\Theta}$ to be the
angle between the polarization of the \taulepton\ and the momentum of
the anti--muon (assuming ${\tau}^{-} {\to} {\mu}^{-} {\mu}^{-}
{\mu}^{+}$) and differentiate models according to their partial decay
width in \cosTheta, having integrated over all other kinematic
variables. Approximately $85\%$ of the \tauleptons\ from electroweak
gauge bosons will be produced in $W$ decays, which result in
characteristic spin patterns. For example, the decay $W^-\to \tau^-
\bar{\nu}_\tau$ produces \tauleptons\ with left--handed helicity only,
so that the polarization vector is antiparallel to the momentum vector
and can thus be determined experimentally. We note that the
polarization of \tauleptons\ from $D$ and $B$ mesons or from $Z$
decays cannot be determined in such simple manner. The \tauleptons\ from
these sources should thus not be included in our polarization
analysis, but instead dilute the signal.

The paper is organized as follows: first, we will present analytic
expressions for the matrix element and differential decay width
derived from an effective Lagrangian with the most general
four--fermion $( {\bar{{\mu}}} {\mu} ) ( {\bar{{\mu}}} {\tau} )$
interactions that do not involve derivatives. The result of this
calculation has been used in a Monte Carlo study to estimate the
detector acceptance for various new physics models. In order to show
how the decay angular distribution can be employed to discriminate
between different models we provide specific predictions for various
supersymmetric, little Higgs and technicolour models, and models with
doubly charged Higgs bosons.  In each case, we will briefly describe
the relevant features of the model, extract the values for the
effective parameters of the general matrix element and differential
decay width expressions, present the dependence of its partial decay
width with respect to \cosTheta\ and the acceptance for the decay
simulated in a typical LHC general purpose detector with the
experimental cuts described above.  Finally, we will briefly discuss
three additional classes of models: those that fit the criterion of
heavy mediating particles, but have no concrete predictions for the
dependence on \cosTheta, those which are very constrained by the
\tautomugamma\ data, and unparticle models.

\section{Model--independent analysis}

We consider an interaction using effective four--fermion vertices.
Derivatives have not been included, since by the equations of motion
these derivatives will be of the order of the lepton masses, which
should be small in comparison to the scale of the new physics.  For
other model--independent investigations based on effective
lepton--flavour--violating interactions, see
Refs.~\cite{Kitano:2000fg}.

The effective Lagrangian is given by (using ${\rho}$ and ${\nu}$ as
Lorentz indices to avoid confusion with the symbol ${\mu}$ being used
for muon spinors)
\begin{eqnarray}\label{eq:l_modind}
{\mathcal{L}} & = & G \Bigg( g_{LL}^{S} ( {\bar{{\mu}}} P_{R} {\mu} ) ( {\bar{{\mu}}} P_{L} {\tau} ) + g_{LR}^{S} ( {\bar{{\mu}}} P_{R} {\mu} ) ( {\bar{{\mu}}} P_{R} {\tau} ) + g_{RL}^{S} ( {\bar{{\mu}}} P_{L} {\mu} ) ( {\bar{{\mu}}} P_{L} {\tau} ) + g_{RR}^{S} ( {\bar{{\mu}}} P_{L} {\mu} ) ( {\bar{{\mu}}} P_{R} {\tau} ) \nonumber\\
 & & \qquad + g_{LL}^{V} ( {\bar{{\mu}}} {\gamma}_{{\nu}} P_{R} {\mu} ) ( {\bar{{\mu}}} {\gamma}^{{\nu}} P_{L} {\tau} ) + g_{LR}^{V} ( {\bar{{\mu}}} {\gamma}_{{\nu}} P_{R} {\mu} ) ( {\bar{{\mu}}} {\gamma}^{{\nu}} P_{R} {\tau} ) \nonumber\\[2mm]
 & & \qquad + g_{RL}^{V} ( {\bar{{\mu}}} {\gamma}_{{\nu}} P_{L} {\mu} ) ( {\bar{{\mu}}} {\gamma}^{{\nu}} P_{L} {\tau} ) + g_{RR}^{V} ( {\bar{{\mu}}} {\gamma}_{{\nu}} P_{L} {\mu} ) ( {\bar{{\mu}}} {\gamma}^{{\nu}} P_{R} {\tau} ) \nonumber\\
 & & \qquad + g_{LR}^{T} \left( {\bar{{\mu}}} {\frac{{{\sigma}_{{\rho}{{\nu}}}}}{{\sqrt{2}}}} P_{R} {\mu} \right) \left( {\bar{{\mu}}} {\frac{{{\sigma}^{{\rho}{{\nu}}}}}{{\sqrt{2}}}} P_{R} {\tau} \right) + g_{RL}^{T} \left( {\bar{{\mu}}} {\frac{{{\sigma}_{{\rho}{{\nu}}}}}{{\sqrt{2}}}} P_{L} {\mu} \right) \left( {\bar{{\mu}}} {\frac{{{\sigma}^{{\rho}{{\nu}}}}}{{\sqrt{2}}}} P_{L} {\tau} \right) \Bigg) \nonumber\\[1mm]
 & {\mydefinedby} & G {\sum_{a, b, c}} g_{ab}^{c} \left( {\bar{{\mu}}} {\Gamma}^{c} {\gamma}^{0} P_{a} {\gamma}^{0} {\mu} \right) \left( {\bar{{\mu}}} {\Gamma}^{c} P_{b} {\tau} \right)\,,
\end{eqnarray}
where $P_{L/R}$ are the left-- and right--handed projection operators,
respectively, with $P_{L/R} = (1 \mp {\gammabar})/2$, and
${\sigma}^{{\rho}{{\nu}}} = i [ {\gamma}^{{\rho}}, {\gamma}^{{\nu}} ]
/ 2$. The symbols $g_{ab}^{c}$ denote the couplings for the various
chiral structures, with $c = S \mbox{ for scalar}, V \mbox{ for
  vector}, T \mbox{ for tensor}$, so that ${\Gamma}^{S} = 1,
{\Gamma}^{V} = {\gamma}^{{\nu}}, {\Gamma}^{T} =
\sigma^{\rho\nu}/\sqrt{2}$, and where $a = \{ L, R \}$ is the
chirality of the anti--muon and $b = \{ L, R \}$ is the chirality of
the \taulepton. [The presence of the ${\gamma}^{0}$ matrices on each
side of $P_{a}$ in the final line is to allow $a$ to be the chirality
of the outgoing anti--muon.]  The overall constant $G$ is
dimensionful, with units of GeV${}^{-2}$.  We will use $G$ to absorb
any normalization of the couplings $g_{ab}^{c}$ for any particular
model, so that we can present the $g_{ab}^{c}$ as integers where
possible.  Note that not all ten couplings are independent, as through
Fierz identities one of the four--fermion terms can be written as a
sum of one or more of the others.  However, we choose to keep the
extra term for convenience.  The use of Fierz identities can also show
that the omitted $g_{LL}^{T} \left( {\bar{{\mu}}}
  {\frac{{{\sigma}_{{\rho}{{\nu}}}}}{{\sqrt{2}}}} P_{R} {\mu} \right)
\left( {\bar{{\mu}}} {\frac{{{\sigma}^{{\rho}{{\nu}}}}}{{\sqrt{2}}}}
  P_{L} {\tau} \right)$ and $g_{RR}^{T} \left( {\bar{{\mu}}}
  {\frac{{{\sigma}_{{\rho}{{\nu}}}}}{{\sqrt{2}}}} P_{L} {\mu} \right)
\left( {\bar{{\mu}}} {\frac{{{\sigma}^{{\rho}{{\nu}}}}}{{\sqrt{2}}}}
  P_{R} {\tau} \right)$ terms are identically zero.

Noting that there is a pair of identical fermions in the final state,
the transition matrix element is given by
\begin{eqnarray}
{\mathcal{M}}& =& G {\sum_{a, b, c}} g_{ab}^{c} \left( {\bar{u}}(
  p_{{{\mu}_{A}}} ) {\Gamma}^{c} {\gamma}^{0} P_{a} {\gamma}^{0} v(
  p_{{\bar{{\mu}}}} ) {\bar{u}}( p_{{{\mu}_{B}}} ) {\Gamma}^{c} P_{b}
  u( p_{{\tau}} )\right. \nonumber\\ && \hphantom{G {\sum_{a, b, c}}
  g_{ab}^{c}}\!\!\! \left. - {\bar{u}}( p_{{{\mu}_{B}}} ) {\Gamma}^{c} {\gamma}^{0} P_{a} {\gamma}^{0} v( p_{{\bar{{\mu}}}} ) {\bar{u}}( p_{{{\mu}_{A}}} ) {\Gamma}^{c} P_{b} u( p_{{\tau}} ) \right)\,,
\end{eqnarray}
where $A$ and $B$ label the muons. While we sum over final--state
spins when squaring the matrix element, we keep the information about
the $\tau$ polarization by using $u( p_{{\tau}} ) {\bar{u}}(
p_{{\tau}} ) = \left( {\slashed{p}}_{{\tau}} + m_{{\tau}} \right)
\left( {\frac{{1 + {\gammabar} {\slashed{n}}}}{2}} \right)$, where
$n^{{\nu}}$ is the polarization vector of the \taulepton.

Squaring this matrix element and performing integrations over the
phase space except for the anti--muon energy and the angle ${\Theta}$
between the polarization of the \taulepton\ and the momentum of the
anti--muon produces the result for the normalized
double--differential decay width:
\begin{equation}
\frac{1}{\Gamma} \frac{\differentiald\Gamma}{\differentiald{x}\,
  {\differentiald}{\cos\Theta}} = {\frac{{6 x^{2} ( a + b\, x + c\, \cos\Theta + d\, x\, {\cos\Theta} )}}{( 4 a + 3 b )}}\,,
\end{equation}
with the approximation that the muons are massless.  Here $x = 2
E_{{\bar{{\mu}}}} / m_{{\tau}}$ is the reduced energy of the
anti--muon, and the coefficients $a, b, c$ and $d$ are given by
\begin{eqnarray}\label{eq:abcd}
a & = & 3 |g_{LL}^{S}|^2 + 12 |g_{LL}^{V}|^2 + 3 |g_{LR}^{S}|^2 + 48 |g_{LR}^{V}|^2 
+ 108 |g_{LR}^{T}|^2 + 3 |g_{RL}^{S}|^2 + 48 |g_{RL}^{V}|^2  + 108 |g_{RL}^{T}|^2 
+ 3 |g_{RR}^{S}|^2   \nonumber\\[1mm] &&
 + 12 |g_{RR}^{V}|^2 - 12\,{\rm Re}\left (g_{LL}^{S} g_{LL}^{V{\ast}} +
g_{RR}^{V} g_{RR}^{S{\ast}}\right) 
- 36\,{\rm Re}\left(g_{LR}^{S} g_{LR}^{T{\ast}} + g_{RL}^{S}
  g_{RL}^{T{\ast}}\right) \,,\nonumber\\[2mm]
b & = & -2 |g_{LL}^{S}|^2 - 8 |g_{LL}^{V}|^2 - 3 |g_{LR}^{S}|^2  - 48
|g_{LR}^{V}|^2 - 108
|g_{LR}^{T}|^2 - 3 |g_{RL}^{S}|^2 - 48 |g_{RL}^{V}|^2 - 108
|g_{RL}^{T}|^2 - 2 |g_{RR}^{S}|^2   \nonumber\\[1mm]
 & &  - 8 |g_{RR}^{V}|^2 + 8\,{\rm Re}\left ( g_{LL}^{S} g_{LL}^{V{\ast}} +
  g_{RR}^{V} g_{RR}^{S{\ast}}\right )
+ 36\,{\rm Re}\left( g_{LR}^{S} g_{LR}^{T{\ast}} + g_{RL}^{S} g_{RL}^{T{\ast}}\right) \,,
  \nonumber\\[2mm]
c & = & |g_{LL}^{S}|^2 + 4
|g_{LL}^{V}|^2 - 3 |g_{LR}^{S}|^2 - 48 |g_{LR}^{V}|^2 - 108
|g_{LR}^{T}|^2 + 3 |g_{RL}^{S}|^2 + 48 |g_{RL}^{V}|^2  + 108
|g_{RL}^{T}|^2 - |g_{RR}^{S}|^2
\nonumber\\[1mm]
 & & - 4
|g_{RR}^{V}|^2  + 36\,{\rm
  Re}\left(g_{LR}^{S} g_{LR}^{T{\ast}}  -  g_{RL}^{S} g_{RL}^{T{\ast}}\right)  -
 4\,{\rm Re}\left( g_{LL}^{S} g_{LL}^{V{\ast}}  -  g_{RR}^{V}
   g_{RR}^{S{\ast}}\right) \,, \nonumber\\[2mm]
d & = & -2 |g_{LL}^{S}|^2 - 8
|g_{LL}^{V}|^2 + 3 |g_{LR}^{S}|^2 + 48 |g_{LR}^{V}|^2 + 108
|g_{LR}^{T}|^2 - 3 |g_{RL}^{S}|^2 - 48 |g_{RL}^{V}|^2  
- 108
|g_{RL}^{T}|^2 + 2 |g_{RR}^{S}|^2 
\nonumber\\[1mm]
 & & + 8
|g_{RR}^{V}|^2 - 36\,
{\rm Re}\left( g_{LR}^{S} g_{LR}^{T{\ast}} - g_{RL}^{S} g_{RL}^{T{\ast}}\right)  + 8\, {\rm Re}\left( g_{LL}^{S} g_{LL}^{V{\ast}}  - g_{RR}^{V} g_{RR}^{S{\ast}}\right)\,.
\end{eqnarray}

Hence the normalized differential decay width in \cosTheta\ is
\begin{equation}\label{eq:ct}
{\frac{1}{{\Gamma}}} {\frac{{{\differentiald}{\Gamma}}}{{{\differentiald}{\cos\Theta}}}} = {\frac{1}{2}} \left( 1 + {\frac{{4 c + 3 d}}{{4 a + 3 b}}} {\cos\Theta} \right)\,,
\end{equation}
where we use ${\Gamma}$ to denote only the width of the \tautothreemu\ 
decay, to save on a proliferation of subscripts, rather than using it
to denote the full ${\tau}$ decay width.  This normalized differential
decay width is now independent of the absolute magnitudes of the
four--fermion couplings. Muon mass effects are included in our full
calculation.  They are suppressed by powers of $m_{{\mu}} /
m_{{\tau}}$ and are thus numerically small.

\section{Model Discrimination}\label{sec:models}

Here we discuss various new physics models and consider how they may
be discriminated using the distribution in \cosTheta.  Specifically,
we consider the $R$--parity--conserving MSSM including see--saw
neutrino masses at large \texttanbeta~\cite{Casas:2001sr, Babu_Kolda},
the $R$--parity--violating MSSM~\cite{RPVMSSM_original}, the Littlest
Higgs model with $T$--parity~\cite{LHT_original}, the
topcolour--assisted technicolour model~\cite{TCtwo_original}, the
Higgs triplet model~\cite{HTM_original}, and the Zee--Babu
model~\cite{ZBM_original}. The decays of \tauleptons\ to ${\mu} {\mu}
{\bar{{\mu}}}$ within each model have been discussed in the
literature, but, as far as we are aware, only to the extent of
predicting total branching ratios in terms of the model parameters,
except for \citereference{LRS_LFV}, which considers forward--backward
asymmetries for the Higgs triplet and Zee--Babu models.

\subsection{Supersymmetric model with see--saw mechanism}

Neutrino masses can be accommodated in the Minimal Supersymmetric
Standard Model (MSSM) by adding singlet chiral superfields that have
right--handed neutrino components and a large Majorana mass term,
leading to the supersymmetric version of the ``see--saw''
mechanism~\cite{see-saw, see-saw_nu_review}.  However, the see--saw
MSSM allows for much stronger influence of the mixing in the neutrino
sector on the charged lepton sector, through large
renormalization--group effects in the slepton
sector~\cite{MSSM_see-saw_renormalization,Hall:1985dx}. Here, we focus
on the potential for a large \tautothreemu\ branching ratio escaping
the \tautomugamma\ bound through an enhanced coupling of the MSSM Higgs
bosons to muons for large \texttanbeta, as proposed
in~\citereference{Babu_Kolda}
\begin{figure} 
\begin{center}
\leavevmode
\includegraphics[width=7cm]{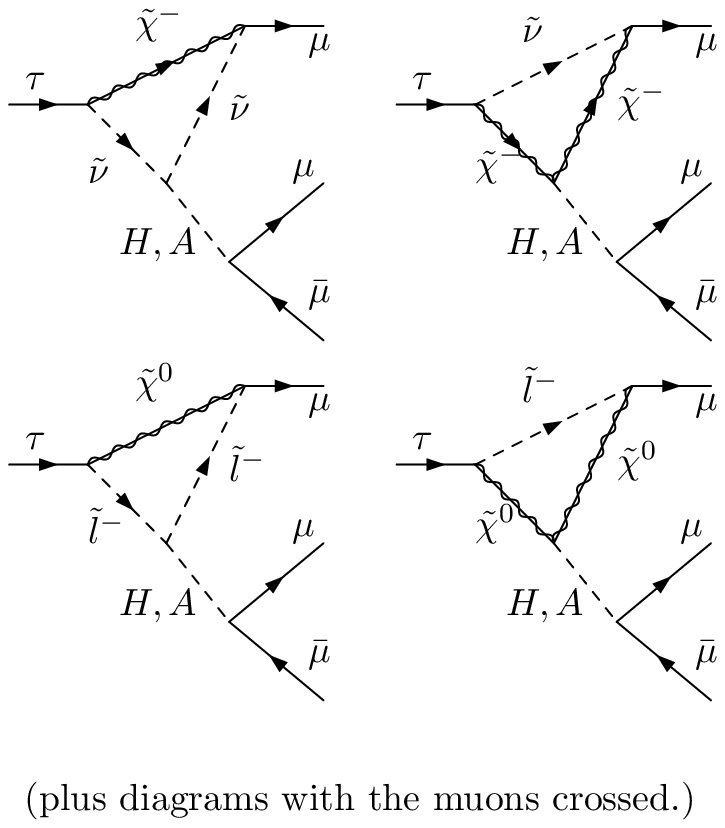}
\caption{Feynman diagrams leading to \tautothreemu\ in the see--saw
  MSSM with large \texttanbeta.}
\label{SSMSSM_tau_to_three_mu_figure}
\end{center}
\end{figure}

We start directly from \citereferencedequation{Babu_Kolda}{15}, which
gives an effective coupling of the MSSM Higgs bosons to a
\taulepton\ and a muon.  We reproduce it here for ease of reference:
\begin{eqnarray}
- {\mathcal{L}} & = & ( 2 G_{F}^{2} )^{1 / 4} {\frac{{m_{{\tau}}
      {\kappa}_{32}}}{{\cos}^{2}{\beta}}} ( {\bar{{\tau}}}_{R}
{\mu}_{L} ) \left[ {\mycos{{{\beta} - {\alpha}}}} h^{0} -
  {\mysin{{{\beta} - {\alpha}}}} H^{0} - i A^{0} \right] + {\rm h.c.} \nonumber\\
 & = & -( 2 G_{F}^{2} )^{1 / 4} m_{{\tau}} {\kappa}_{32} (
 {\tan\beta} )^{2} ( {\bar{{\tau}}} P_{L} {\mu} ) [ H^{0} + i
 A^{0} ] + {\rm h.c.} \nonumber\\
 & = & -( 2 G_{F}^{2} )^{1 / 4} m_{{\tau}} ( {\tan\beta} )^{2} \left[ {\kappa}_{32} ( {\bar{{\tau}}} P_{L} {\mu} ) [ H^{0} + i A^{0} ] + {\kappa}_{32}^{{\ast}} ( {\bar{{\mu}}} P_{R} {\tau} ) [ H^{0} - i A^{0} ] \right]\,,
\end{eqnarray}
where in the second line we have explicitly taken the large $m_{A}$
(${\alpha} {\to} {\beta} - {\pi} / 2$) and large \texttanbeta\ limits.
The flavour--diagonal muon coupling to MSSM Higgs
bosons~\cite{Rosiek:1995kg} in these limits is given by
\begin{equation}
{\mathcal{L}}_{{{\mu}-\text{Higgs}}} = {\frac{e}{\sin\theta_{W}}} {\frac{{m_{{\mu}}}}{{2 m_{W}}}} {\bar{{\mu}}} \left[ h^{0} + {\tan\beta} H^{0} - i {\tan\beta} {\gammabar} A^{0} \right] {\mu}\,,
\end{equation}
so integrating out the $H^{0}$ and $A^{0}$ fields leads to the
relevant part of the effective Lagrangian given by
\begin{eqnarray}\label{eq:leff_ssmssm}
{\mathcal{L}}_{{\text{eff}}} & = & -( 2 G_{F}^{2} )^{1 / 4} m_{{\tau}}
{\kappa}_{32}^{{\ast}} ( {\tan\beta} )^{2} 
{\frac{e}{{\sin\theta_{W}}}} {\frac{{m_{{\mu}}}}{{2 m_{W}}}} ( {\bar{{\mu}}} P_{R} {\tau} ) {\bar{{\mu}}} \left[ {\frac{1}{{m_{{H^{0}}}^{2}}}} {\tan\beta} - {\tan\beta} {\gammabar} {\frac{1}{{m_{A}^{2}}}} \right] {\mu} \nonumber\\
 & = & -( 2 G_{F}^{2} )^{1 / 4} m_{{\tau}} {\kappa}_{32}^{{\ast}} ( {\tan\beta} )^{3} {\frac{e}{{\sin\theta_{W}}}} {\frac{{m_{{\mu}}}}{{2 m_{W}}}} {\frac{2}{{m_{A}^{2}}}} ( {\bar{{\mu}}} P_{L} {\mu} ) ( {\bar{{\mu}}} P_{R} {\tau} )\,.
\end{eqnarray}
In the second line of (\ref{eq:leff_ssmssm}) we have used the fact
that in the large $m_{A}$ limit, the mass of the heavier $CP$--even
Higgs boson, $H^{0}$, is approximately equal to $m_{A}$.  We can
easily see that the only non--zero effective coupling $g_{ab}^{c}$ in
the Lagrangian (\ref{eq:l_modind}) is $g_{RR}^{S}$, which we set to
$1$.

Performing the phase--space integrations analytically with massless muons gives the result
\begin{equation}\label{eq:mssm}
{\frac{1}{{\Gamma}}}
{\frac{{{\differentiald}{\Gamma}}}{{{\differentiald} {\cos\Theta}
      }}} = \frac{3 + 
\cos\Theta}{6}\,, 
\end{equation}
which can be derived from Eq.~(\ref{eq:abcd}) and (\ref{eq:ct}).  In
order to quantify the effect of muon masses, we display
Eq.(\ref{eq:mssm}) in \citefigure{SSMSSM_costheta_plot}, along with
the result of a numerical calculation incorporating massive muons. As
expected, the muon mass effects are very small.  A Monte--Carlo
simulation of the decay \tautothreemu\ within the MSSM see--saw model
for a typical LHC general purpose detector (with cuts as described in
Section~\ref{sec:intro}) shows an acceptance of $30\%$.

\begin{figure}
\vspace*{-10mm}
\begin{center}
\includegraphics[width=12cm]{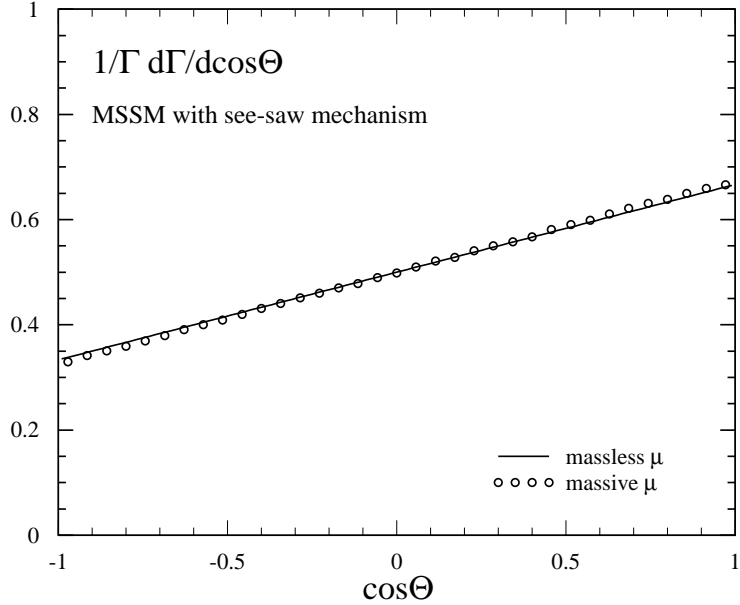}
\caption{$1/\Gamma$ $\differentiald\Gamma/\differentiald\cos\Theta$
  for the supersymmetric see--saw model with large \texttanbeta\ 
  defined in~\citereference{Babu_Kolda}. We display the analytic
  result obtained for massless muons together with the numerical
  calculation including all mass effects.}
\label{SSMSSM_costheta_plot}
\end{center}
\end{figure}

\subsection{Supersymmetric model with $R$--parity violation}

The MSSM without $R$--parity (see \eg~Refs.~\cite{RPV_review} for
reviews) includes an additional set of Yukawa interactions, between
the charged leptons and sneutrinos, which are not necessarily
diagonalized when the Higgs--lepton Yukawa interactions are
diagonalized, allowing for the possibility of
charged--lepton--flavour--violation by a scalar particle at
tree--level. The $R$--parity--violating part of the superpotential is
given by
\begin{equation}\label{RPV_superpotential_equation}
W_{{\Slashed{R}}_{p}} = 
        \epsilon_{ab}\left(\frac{1}{2}\lambda_{ijk} L_i^aL_j^b{\bar E}_k
        + \lambda_{ijk}^\prime L_i^aQ_j^b{\bar D}_k
        + \kappa_i L_i^aH_u^b\right) 
+ \frac{1}{2} \epsilon_{rst}\lambda_{ijk}^
{\prime\prime}{\bar U}_i^r{\bar D}_j^s {\bar D}_k^t \,.  
\end{equation}
Here, $i, j, k = 1, 2, 3$ are generation indices, $a, b = 1, 2$ are $SU(2)$ and
$r, s, t = 1, 2, 3$ are $SU(3)$ indices. $L, \bar{E}$ denote the lepton doublet
and singlet left--chiral superfields; $Q,\bar{U}, \bar{D}$ denote the
quark doublet and singlet superfields, respectively.
$\lambda, \lambda', \lambda''$ are dimensionless coupling constants and
$\kappa$ is a mass mixing parameter.

The terms relevant to the decay \tautothreemu\ are
$\epsilon_{ab}\frac{1}{2}\lambda_{ijk} L_i^aL_j^b{\bar E}_k$
  plus its Hermitian conjugate, which give interaction terms
  ${\lambda}_{i23} {\superpartner{{\nu}}}_{iL} {\bar{{\mu}}} P_{R}
  {\tau}$, ${\lambda}_{i32}^{{\ast}}
  {\superpartner{{\nu}}}_{iL}^{{\ast}} {\bar{{\mu}}} P_{L} {\tau}$,
  ${\lambda}_{i22} {\superpartner{{\nu}}}_{iL} {\bar{{\mu}}} P_{R}
  {\mu}$ and ${\lambda}_{i22}^{{\ast}}
  {\superpartner{{\nu}}}_{iL}^{{\ast}} {\bar{{\mu}}} P_{L} {\mu}$.
  These lead to the \tautothreemu\ diagrams shown
  in~\citefigure{RPVMSSM_tau_to_three_mu_figure}.
\begin{figure} 
\begin{center}
\leavevmode
\includegraphics[width=7cm]{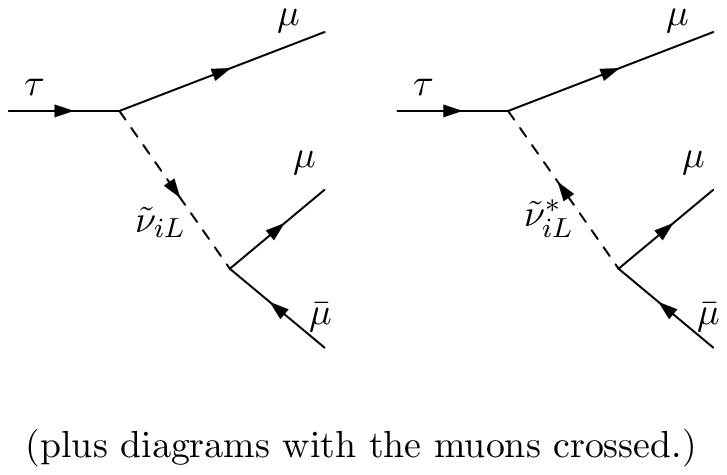}
\caption{Feynman diagrams leading to \tautothreemu\ in the MSSM without $R$--parity.}
\label{RPVMSSM_tau_to_three_mu_figure}
\end{center}
\end{figure}
Integrating out the sneutrinos, which are constrained to be heavy
($m_{{\superpartner{{\nu}}}} \gg m_{{\tau}}$), gives the following
terms in the effective Lagrangian:
\begin{equation}
{\mathcal{L}}_{{\text{eff}}} = {\sum_{i}} \left( {\frac{1}{{m_{{{\superpartner{{\nu}}}_{iL}}}^{2}}}} {\lambda}_{i32} {\lambda}_{i22}^{{\ast}} ( {\bar{{\mu}}} P_{R} {\mu} ) ( {\bar{{\mu}}} P_{L} {\tau} ) + {\frac{1}{{m_{{{\superpartner{{\nu}}}_{iL}}}^{2}}}} {\lambda}_{i22} {\lambda}_{i23}^{{\ast}} ( {\bar{{\mu}}} P_{L} {\mu} ) ( {\bar{{\mu}}} P_{R} {\tau} ) \right)\,,
\end{equation}
from which we can read off that $g_{LL}^{S}$ is proportional to
${\lambda}_{i32} {\lambda}_{i22}^{{\ast}}$ and $g_{RR}^{S}$ is
proportional to ${\lambda}_{i22} {\lambda}_{i23}^{{\ast}}$, $g_{LR}^{S} = g_{RL}^{S} = 0$, and all the vector and tensor couplings
are also zero.

Bottom--up approaches to constraining the MSSM without $R$--parity
generally consider the minimal number of couplings to be non--zero for
any individual constraining process~\cite{RPV_constraint_papers},
\eg~constraints from the current upper bound on \tautothreemu\ assume
either ${\lambda}_{i32} {\lambda}_{i22}^{{\ast}} \neq 0$ and
${\lambda}_{i22} {\lambda}_{i23}^{{\ast}} = 0$ or the other way
around, to obtain a conservative upper bound in the absence of
destructive interference~\cite{HD_MK_BOL_RPV_bounds}. Here we take two
benchmark scenarios: one where $g_{LL}^{S} = 1$ and $g_{RR}^{S} = 0$,
designated ``L'', and the other where $g_{LL}^{S} = 0$ and $g_{RR}^{S}
= 1$, designated ``R''.  If the decay \tautothreemu\ is measurable at
the LHC because this model is correct, then it is almost certain that
it will have been established from other signals, and the measurement
distinguishing between the two benchmarks would be most useful for
measuring the $R$--parity--violating couplings.

Performing the phase--space integrations analytically with massless
muons gives the result
\begin{equation}
{\frac{1}{{\Gamma}}} {\frac{{{\differentiald}{\Gamma}}}{{{\differentiald} {\cos\Theta} }}} = \left\{ \begin{array}{l l}\begin{displaystyle}
{\frac{{3 - {\cos\Theta}}}{6}} \end{displaystyle} & \text{ for L:}\; g_{LL}^{S} = 1,
g_{RR}^{S} = 0 \,, \\[4mm]
\begin{displaystyle}
{\frac{3 + \cos\Theta}{6}} \end{displaystyle} & \text{ for R:}\; g_{LL}^{S} = 0,
g_{RR}^{S} = 1\,.
\end{array} \right.
\end{equation}
Simulation of these decays for a typical LHC general purpose detector
shows an acceptance of $25\%$ and $30\%$ for ``L'' and ``R'',
respectively.

\subsection{Littlest Higgs model with $T$--parity}

The additional gauge group(s) and fermion multiplets in the Littlest
Higgs model with $T$--parity (LHT model) (see \eg~Refs.~\cite{LHT_review}) allow
for flavour--changing interactions through loops of $T$--odd particles
with potentially very different PMNS-- or CKM--like matrices.  We
focus on the way this allows for \tautothreemu\ decays as discussed
in~\citereference{Buras_LHT_LFV}.

\begin{figure} 
\begin{center}
\leavevmode
\includegraphics[width=7cm]{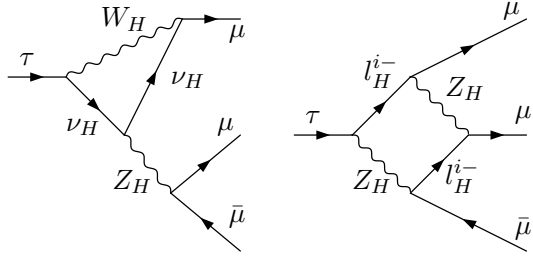}
\caption{Example Feynman diagrams leading to \tautothreemu\ in the
  Littlest Higgs model with $T$--parity.  For the full set, see
  \citereference{Buras_LHT_LFV}.}
\label{LHT_tau_to_three_mu_figure}
\end{center}
\end{figure}

We start directly from
\citethreereferencedequations{Buras_LHT_LFV}{5.2}{5.3}{5.4}, which
give four--point amplitudes for ${\tau} {\to} \mu \mu {\bar{\mu}}$ in the
LHT.  We reproduce them here for ease of reference:
\begin{eqnarray}
{\mathcal{A}}_{{{\gamma}'}} & = & {\frac{{G_{F}}}{{\sqrt{2}}}}
{\frac{{e^{2}}}{{8 {\pi}^{2}}}} {\frac{1}{{q^{2}}}}
{\bar{D}}_{{\text{odd}}}^{{\prime}{\tau}\mu} [ {\bar{\mu}}( p_{1} ) (
m_{{\tau}} i {\sigma}_{{\alpha}{\beta}} q^{{\beta}} ( 1 + {\gamma}_{5}
) ) {\tau}( p ) ] [ {\bar{\mu}}( p_{2} ) {\gamma}^{{\alpha}} \mu(
p_{3} )] - ( p_{1} {\leftrightarrow} p_{2} )\,, \\
{\mathcal{A}}_{{\gamma}} & = & -\left[ 4 {\frac{{G_{F}}}{{\sqrt{2}}}}
  {\frac{{e^{2}}}{{8 {\pi}^{2}}}}
  {\bar{Z}}_{{\text{odd}}}^{{\prime}{\tau}\mu} \left[ {\bar{\mu}}(
    p_{1} ) {\gamma}_{{\alpha}} ( 1 - {\gamma}_{5} ) {\tau}( p )
  \right] \left[ {\bar{\mu}}( p_{2} ) {\gamma}^{{\alpha}} \mu( p_{3} )
  \right] - ( p_{1} {\leftrightarrow} p_{2} ) \right] \,, \\
{\mathcal{A}}_{{\text{box}}} & = & 2 {\frac{{G_{F}}}{{\sqrt{2}}}} {\frac{{\alpha}}{{2 {\pi} {\sin}^{2} {\theta}_{W}}}} {\bar{Y}}_{\mu,{\text{odd}}}^{{\prime}{\tau}\mu} \left[ {\bar{\mu}}( p_{1} ) {\gamma}_{{\alpha}} ( 1 - {\gamma}_{5} ) {\tau}( p ) \right] \left[ {\bar{\mu}}( p_{2} ) {\gamma}^{{\alpha}} ( 1 - {\gamma}_{5} ) \mu( p_{3} ) \right]\,.
\end{eqnarray}
The structure of the amplitude ${\mathcal{A}}_{{\gamma}}$ corresponds
to setting the couplings $g_{LL}^{V} = g_{RL}^{V} = 1$ (and all others
to zero) in the general effective Lagrangian (\ref{eq:l_modind}),
while the structure of the amplitude ${\mathcal{A}}_{{\text{box}}}$
corresponds to setting $g_{RL}^{V} = 1$ and all other couplings to
zero.  The amplitude ${\mathcal{A}}_{{{\gamma}'}}$ is already tightly
constrained by \tautomugamma, so we ignore it in the following.  [We
also found for input mirror--lepton to mirror--$W$ boson mass ratios in
the range from $0.1$ to $10$ that ${\mathcal{A}}_{{{\gamma}'}}$ is
much smaller than the other amplitudes.]  The functions
${\bar{Z}}_{{\text{odd}}}^{{\prime}{\tau}{\mu}}$ and
${\bar{Y}}_{{\mu},{\text{odd}}}^{{\prime}{\tau}{\mu}}$, defined
in~\citereference{Buras_LHT_LFV}, encode all the information from the
$T$--odd particles in the decay, and can vary strongly in their
relative magnitude. However, we found that for mirror--lepton to
mirror--$W$ mass ratios below $2$,
${\bar{Z}}_{{\text{odd}}}^{{\prime}{\tau}{\mu}}\, {\ll}\,
{\bar{Y}}_{{\mu},{\text{odd}}}^{{\prime}{\tau}{\mu}}$ while for ratios
above about $4$ ${\bar{Z}}_{{\text{odd}}}^{{\prime}{\tau}{\mu}}$
becomes dominant.  As a pair of benchmarks, we thus take
${\bar{Z}}_{{\text{odd}}}^{{\prime}{\tau}{\mu}} \,{\gg}\,
{\bar{Y}}_{{\mu},{\text{odd}}}^{{\prime}{\tau}{\mu}}$, so $g_{LL}^{V}
= g_{RL}^{V} = 1$ (and all other couplings set to zero), designated
``Z'', and ${\bar{Z}}_{{\text{odd}}}^{{\prime}{\tau}{\mu}} \,{\ll}\,
{\bar{Y}}_{{\mu},{\text{odd}}}^{{\prime}{\tau}{\mu}}$, so $g_{RL}^{V}
= 1$ and all others zero, designated ``Y''.

Performing the phase--space integrations analytically with massless
muons gives the result
\begin{equation}
{\frac{1}{{\Gamma}}} {\frac{{{\differentiald}{\Gamma}}}{{{\differentiald} {\cos\Theta} }}} = \left\{ \begin{array}{l l}
\begin{displaystyle} {\frac{{9 + 5 {\cos\Theta}}}{18}}
\end{displaystyle}& \text{ for Z:}\; g_{LL}^{V} = 1, g_{RL}^{V} = 1
\,, \\[4mm]
\begin{displaystyle} {\frac{{1 + {\cos\Theta}}}{2}}
\end{displaystyle}   & \text{ for Y:}\; g_{LL}^{V} = 0, g_{RL}^{V} = 1\,.
\end{array} \right.
\end{equation}
Simulation of these decays in the LHC environment show an acceptance
of $27\%$ for both the scenarios ``Z'' and ``Y''.

\subsection{Topcolour--assisted technicolour model}

Topcolour--assisted technicolour models~\cite{TCtwo_original}, are
based on the product of gauge groups $SU( 3 )_{1} {\times}SU( 3
)_{2} {\times}$ $U( 1 )_{Y1} {\times} U( 1 )_{Y2} {\times} SU( 2 )_{L}$
which is broken to $SU( 3 )_{{\text{QCD}}} {\times} U( 1
)_{{\text{EM}}}$.  The group $SU( 3 )_{2} {\times} U( 1 )_{Y2}$
couples preferentially to the lighter two generations, while $SU( 3
)_{1} {\times} U( 1 )_{Y1}$ couples preferentially to the heaviest
generation.  The generation--dependence of the couplings lead in
general to flavour--non--diagonal couplings of the heavy ${Z'}$ vector
gauge boson associated with the broken $U( 1 )$ gauge group once the
fermions have been diagonalized to the mass--eigenstate basis.

The top--pion, ${\pi}_{t}$, a condensate which breaks the $SU( 2
)_{L}$ vacuum, can also mediate flavour--changing interactions.  Since
both the ${Z'}$ and the top--pion masses and couplings are unknowns,
we consider the two extreme cases where either of the two particles
dominates the decay: we designate the ${Z'}$ case by ``Z'' and the
top--pion case by ``P'', which occur for example when $m_{\pi_t} \gg
m_{Z'}$ or $m_{\pi_t} \ll m_{Z'}$, respectively.

\begin{figure} 
\begin{center}
\leavevmode
\includegraphics[width=7cm]{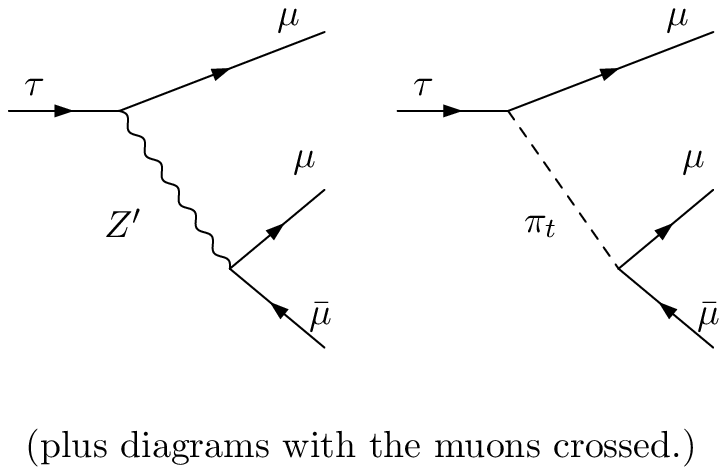}
\caption{Feynman diagrams leading to \tautothreemu\ in the topcolour--assisted technicolour model.}
\label{TCtwo_tau_to_three_mu_figure}
\end{center}
\end{figure}

We consider \tautothreemu\ through the heavy ${Z'}$, using couplings as
given in~\citereference{TCtwo_Z_LFV}.  The flavour--diagonal couplings
read
\begin{eqnarray}
{\mathcal{L}}_{{Z'}}^{{\text{FD}}} & = & -{\frac{1}{2}} g_{1} {\cot} {{\theta}'} {Z'}_{{\mu}} ( {\bar{{\tau}}}_{L} {\gamma}^{{\mu}} {\tau}_{L} + 2 {\bar{{\tau}}}_{R} {\gamma}^{{\mu}} {\tau}_{R} ) \nonumber\\
 & & - {\frac{1}{2}} g_{1} {\tan} {{\theta}'} {Z'}_{{\mu}} ( {\bar{{\mu}}}_{L} {\gamma}^{{\mu}} {\mu}_{L} + 2 {\bar{{\mu}}}_{R} {\gamma}^{{\mu}} {\mu}_{R} + {\bar{e}}_{L} {\gamma}^{{\mu}} e_{L} + 2 {\bar{e}}_{R} {\gamma}^{{\mu}} e_{R} )\,,
\end{eqnarray}
while the flavour--changing couplings are given by
\begin{eqnarray}
{\mathcal{L}}_{{Z'}}^{{\text{FC}}} & = & -{\frac{1}{2}} g_{1} {Z'}_{{\mu}} \left[ k_{{\tau}{\mu}} ( {\bar{{\tau}}}_{L} {\gamma}^{{\mu}} {\mu}_{L} + 2 {\bar{{\tau}}}_{R} {\gamma}^{{\mu}} {\mu}_{R} ) + k_{{\tau}e} ( {\bar{{\tau}}}_{L} {\gamma}^{{\mu}} e_{L} + 2 {\bar{{\tau}}}_{R} {\gamma}^{{\mu}} e_{R} ) \right. \nonumber\\
 & & + \left. k_{{\mu}e} {\tan}^{2} {{\theta}'} {Z'}_{{\mu}} (
   {\bar{{\mu}}}_{L} {\gamma}^{{\mu}} {\mu}_{L} + 2 {\bar{{\mu}}}_{R}
   {\gamma}^{{\mu}} {\mu}_{R} + {\bar{e}}_{L} {\gamma}^{{\mu}} e_{L} +
   2 {\bar{e}}_{R} {\gamma}^{{\mu}} e_{R} )  \right] + {\rm h.c.}\,,
\end{eqnarray}
where $g_{1} = {\sqrt{4 {\pi} {\alpha}}} / {\cos\theta_{W}}$
and ${{\theta}'}$ is the mixing angle\footnote{We note that, in
  \citereference{TCtwo_Z_LFV}, the formula for the flavour--changing
  terms actually has ${\tan}^{2} {\theta}$ rather than ${\tan}^{2}
  {{\theta}'}$, but we assume that this was a typographical error.  We
  also assume that the Hermitian conjugate was also meant to be
  present in the equation.  However, we disagree with
  \citereferencedequation{TCtwo_Z_LFV}{4}, where we believe that the
  numerical factor should be 7/4096 rather than 25/384 for the case of
  identical particles in the final state (${\tau} {\to} {\mu} {\mu}
  {\bar{{\mu}}}, {\mu} {\mu} {\bar{e}}, e e {\bar{{\mu}}}, e e
  {\bar{e}}$) and 25/1536 for the case of no identical particles in
  the final state (${\tau} {\to} {\mu} e {\bar{{\mu}}}, {\mu} e
  {\bar{e}}$).} for the heavy ${Z'}$.  The $k_{{l^{i}}{l^{j}}}$ are
flavour mixing factors.

Integrating out the ${Z'}$ gives the following relevant terms in the
effective Lagrangian:
\begin{eqnarray}
{\mathcal{L}}_{{\text{eff}}} & = & {\frac{1}{{m_{{Z'}}^{2}}}}
{\frac{{{\pi} {\alpha}}}{{( {\cos\theta_{W}} )^{2}}}} 
{\tan\theta'} k_{{\tau}{\mu}} \left( {\bar{{\mu}}} {\gamma}^{{\mu}} P_{L} {\mu} {\bar{{\mu}}} {\gamma}^{{\mu}} P_{L} {\tau} + 2 {\bar{{\mu}}} {\gamma}^{{\mu}} P_{R} {\mu} {\bar{{\mu}}} {\gamma}^{{\mu}} P_{L} {\tau} \right. \nonumber\\[1mm]
 & & \left. + 2 {\bar{{\mu}}} {\gamma}^{{\mu}} P_{L} {\mu} {\bar{{\mu}}} {\gamma}^{{\mu}} P_{R} {\tau} + 4 {\bar{{\mu}}} {\gamma}^{{\mu}} P_{R} {\mu} {\bar{{\mu}}} {\gamma}^{{\mu}} P_{R} {\tau} \right)\,,
\end{eqnarray}
from which we can read off that $g_{RL}^{V} = 1$, $g_{LL}^{V} =
g_{RR}^{V} = 2$, $g_{LR}^{V} = 4$ and all other couplings are zero.

Now we consider \tautothreemu\ through the ${\pi}_{t}$, using couplings
as given in~\citereference{TCtwo_pi_LFV}.  The couplings to the
top--pion are given by the following terms in the Lagrangian:
\begin{eqnarray}
{\mathcal{L}}_{{{\pi}_{t}}}^{{\text{FD}}} & = & \left[ {\frac{{m_{t}}}{{{\sqrt{2}} F_{t}}}} {\frac{{\sqrt{{{\nu}_{{\omega}}^{2} - F_{t}^{2}}}}}{{{\nu}_{{\omega}}}}} [ K_{UR}^{tt} K_{UL}^{tt{\ast}} {\bar{t}} {\gammabar} t {\pi}_{t}^{0} + {\frac{{m_{b} - m_{b}^{{\prime}}}}{{m_{t}}}} {\bar{b}} {\gammabar} b {\pi}_{t}^{0} + K_{UR}^{tc} K_{UL}^{tt{\ast}} {\bar{t}}_{L} {\gammabar} c_{R} {\pi}_{t}^{0} ] \right. \nonumber\\
 & & \left. + {\frac{{m_{l}}}{{{\sqrt{2}} {\nu}_{{\omega}}}}}
   {\bar{l}} {\gammabar} l {\pi}_{t}^{0} + {\frac{{m_{l}}}{{{\sqrt{2}}
         {\nu}_{{\omega}}}}} K_{{\tau}i} {\bar{{\tau}}} {\gammabar}
   l_{i} {\pi}_{t}^{0} \right] + {\rm h.c.}\,, 
\end{eqnarray}
where $m_b^\prime \sim 0.1\, \epsilon\, m_t$ is the part of the bottom
quark mass generated by extended technicolour ($\epsilon \ll 1$ is a
small parameter).  The symbol $K$ denotes flavour mixing factors,
$\nu_\omega = v/\sqrt{2}$ where $v \simeq 246$~GeV is the scale of
electroweak symmetry breaking and $F_t$ is the top--pion decay
constant.

Integrating out the ${\pi}_{t}$ gives the following relevant terms in
the effective Lagrangian:
\begin{eqnarray}
{\mathcal{L}}_{{\text{eff}}} & = & -{\frac{{m_{l} m_{{l_{j}}}}}{{2 {\nu}_{{\omega}}^{2}}}} K_{{\tau}i}^{{\ast}} {\bar{{l_{j}}}} {\gammabar} l_{j} {\bar{{l_{i}}}} {\gammabar} {\tau} \nonumber\\
 & = & -{\frac{{m_{l} m_{{l_{j}}}}}{{2 {\nu}_{{\omega}}^{2}}}} K_{{\tau}i}^{{\ast}} {\bar{{l_{j}}}} ( P_{R} - P_{L} ) l_{j} {\bar{{l_{i}}}} ( P_{R} - P_{L} ) {\tau}\,,
\end{eqnarray}
from which we can read off that $g_{LL}^{S} = g_{RR}^{S} = 1$,
$g_{RL}^{S} = g_{LR}^{S} = -1$, and all others are zero.

Performing the phase--space integrations analytically with massless
muons gives the result
\begin{equation}
{\frac{1}{{\Gamma}}} {\frac{{{\differentiald}{\Gamma}}}{{{\differentiald} {\cos\Theta} }}} = \left\{ \begin{array}{l l}
\begin{displaystyle}{\frac{{7 - 5
        {\cos\Theta}}}{14}}\end{displaystyle} & \text{ for Z:}\;
g_{RL}^{V} = 1, g_{LL}^{V} = g_{RR}^{V} = 2, g_{LR}^{V} = 4 \,, \\
\begin{displaystyle}{\frac{{1}}{2}}\end{displaystyle}
& \text{ for P:}\; g_{LL}^{S} = g_{RR}^{S} = 1, g_{RL}^{S} =
g_{LR}^{S} = -1\,.
\end{array} \right.
\end{equation}

Simulation of \tautothreemu\ decays within the topcolour--assisted
technicolour model show an acceptance of $29\%$ and $28\%$ for the
``Z'' and ``P'' scenarios, respectively.

\subsection{Models with doubly--charged Higgs bosons}

Models with doubly--charged Higgs bosons can mediate LFV $\tau$ decays
through Feynman diagrams like that depicted in
\citefigure{HTM_tau_to_three_mu_figure}. To be specific, we discuss
two concrete models, the Higgs Triplet model~\cite{HTM_original}, and
the Zee--Babu model~\cite{ZBM_original}, both of which have been
discussed in the context of LFV $\tau$ decays previously in
\citereference{LRS_LFV}.

In the Higgs--triplet model neutrinos are given Majorana masses through
the addition of a triplet of $SU(2)$ with hypercharge $2$ (no
right--handed neutrinos are introduced) which obtains a vacuum
expectation value. The doubly--charged components of this triplet can
also mediate the decay \tautothreemu.

The Zee--Babu model~\cite{ZBM_original} is another model in which
neutrinos are given Majorana masses without the introduction of
right--handed neutrinos, through the introduction of additional
singly-- and doubly--charged scalars which are singlets of $SU(2)$.
The neutrino masses are generated radiatively at the two--loop level.

\begin{figure} 
\begin{center}
\leavevmode
\includegraphics[width=4cm]{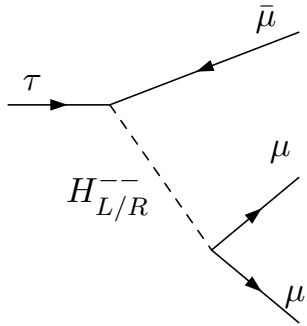}
\caption{Feynman diagram leading to \tautothreemu\ in 
  models with doubly--charged Higgs bosons.}
\label{HTM_tau_to_three_mu_figure}
\end{center}
\end{figure}

We start our discussion from \citereferencedequation{LRS_LFV}{61}
combined with Table 2 therein.  We reproduce the (Hermitian conjugate
of the) equation and the relevant information from the table here for
ease of reference:
\begin{equation}\label{eq:doubleh}
{\mathcal{L}}_{{\text{eff}}} = \left\{ {\frac{{-4 G_{F}}}{{\sqrt{2}}}} g_{3}^{{\ast}} ( {\bar{{\mu}}} {\gamma}^{{\nu}} P_{R} {\mu} ) ( {\bar{{\mu}}} {\gamma}_{{\nu}} P_{R} {\tau} ) + g_{4}^{{\ast}} ( {\bar{{\mu}}} {\gamma}^{{\nu}} P_{L} {\mu} ) ( {\bar{{\mu}}} {\gamma}_{{\nu}} P_{L} {\tau} ) \right\}\,,
\end{equation}
where
\begin{equation}\label{eq:ht}
{\frac{{-4 G_{F}}}{{\sqrt{2}}}} g_{3} =  0 \qquad  \mbox{and} \qquad
{\frac{{-4 G_{F}}}{{\sqrt{2}}}} g_{4} =  {\frac{{h_{{\mu}{\mu}} h_{{\tau}{\mu}}^{{\ast}}}}{{M_{{H_{L}^{{\pm}{\pm}}}}^{2}}}} 
\end{equation}
for the Higgs triplet model, and 
\begin{equation}\label{eq:zee}
{\frac{{-4 G_{F}}}{{\sqrt{2}}}} g_{3} =  {\frac{{h_{{\mu}{\mu}} h_{{\tau}{\mu}}^{{\ast}}}}{{M_{{H_{R}^{{\pm}{\pm}}}}^{2}}}} \qquad \mbox{and} \qquad
{\frac{{-4 G_{F}}}{{\sqrt{2}}}} g_{4} = 0
\end{equation}
for the Zee--Babu model. The constants $h$ are mixing factors and
$M_{{H_{L/R}^{{\pm}{\pm}}}}$ is the mass of the doubly--charged scalar
boson mediating the decay.

The structure of Eq.~(\ref{eq:doubleh}) to (\ref{eq:zee}) corresponds
to setting $g_{RL}^{V} = 1$ and all other couplings to zero for the
Higgs triplet model, while the $g_{LR}^{V} = 1$ is the only non--zero
coupling in the Zee--Babu model. We note that LFV $\tau$ decays with a
chirality structure analogous to that of the Higgs triplet model are
predicted in seesaw models with scalar triplets, see
\eg~Ref.~\cite{Abada:2007ux}.

Performing the phase--space integrations analytically with massless
muons gives the result
\begin{equation}
{\frac{1}{{\Gamma}}} {\frac{{{\differentiald}{\Gamma}}}{{{\differentiald} {\cos\Theta} }}} 
= \left\{ \begin{array}{l l}
\begin{displaystyle}{\frac{{1 + {\cos\Theta}}}{2}}\end{displaystyle} & 
\text{ for the Higgs triplet model:}\; g_{RL}^{V} = 1 \,, \\[4mm]
\begin{displaystyle}{\frac{{1 - {\cos\Theta}}}{2}}\end{displaystyle}                         & 
\text{ for the Zee--Babu model:}\; g_{LR}^{V} = 1\,.
\end{array} \right.
\end{equation}
Simulation of this decay for a typical LHC general purpose detector
shows an acceptance of $25\%$ and $30\%$ for the Higgs triplet and
Zee--Babu models, respectively.

\subsection{Other Models}

Finally, let us very briefly discuss some other new physics models,
namely those that fit the criterion of heavy mediating particles, but
have no concrete predictions for the dependence on \cosTheta, those
which are very constrained by the \tautomugamma\ data, and unparticle
models.

\subsubsection{Heavy--particle--mediated}

Left--Right Symmetric models~\cite{LRS_original} introduce a second
$SU( 2 )$ gauge group for the right--handed fermions, arranging them
in appropriate doublets.  The scalar sector is more complex than that
of the SM, consisting of a Higgs bi--doublet and a Higgs triplet for
each gauge group.  The doubly--charged components of the triplets can
mediate lepton flavour violation~\cite{LRS_LFV}, but we find that the
model is not predictive enough to meaningfully differentiate it
through ${\Theta}$: the slope of $1 / {\Gamma}$ 
${\differentiald}{\Gamma} / {\differentiald} {\cos\Theta}$
depends on which $H^{{\pm}{\pm}}$ dominates the decay and on the
mixing matrices, which are entirely unknown in the right--handed
sector.

\subsubsection{Photon--mediated}

Any model that predicts lepton flavour violation through loops will
predict \tautothreemu\ through a virtual photon.  However, unless the
model predicts other processes that can facilitate \tautothreemu\ (such
as the see--saw MSSM), the bounds from \tautomugamma\ mean that
\tautothreemu\ from such models can barely be observed at LHC, let
alone with sufficient statistics to discriminate models through
angular distributions.

\subsubsection{Unparticle--mediated}

The phenomenology of ``unparticles''~\cite{unparticle_original} has
recently attracted a considerable amount of attention. In particular,
\citetworeferences{unparticle_LFV_first}{unparticle_LFV_II}\ consider
the effects of unparticles mediating lepton flavour violation.
However, the hypothesized unparticle sector has no predicted form for
the couplings to the SM particles, and so has no unique effect on
${\differentiald}{\Gamma} / {\differentiald} {\cos\Theta}$.
The unparticle nature of \tautothreemu\ would instead show up in the
differential cross--section with respect to the energies of the muons,
analogously to the analysis of the electron energies in
\citereference{unparticle_LFV_II}.

\subsection{Discrimination potential at the LHC}

\begin{figure}
\vspace*{-10mm}
\begin{center}
\includegraphics[width=13cm]{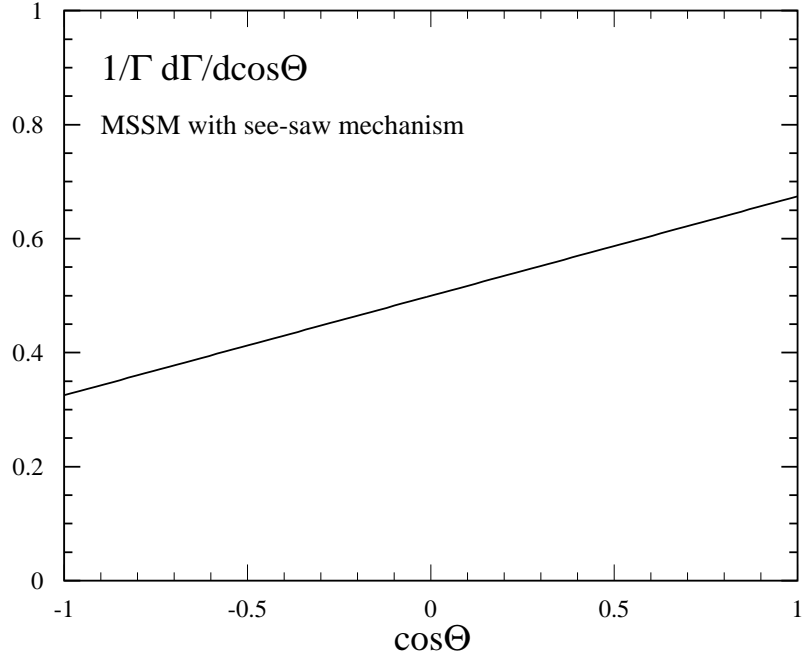}\\[-5mm]
\includegraphics[width=13cm]{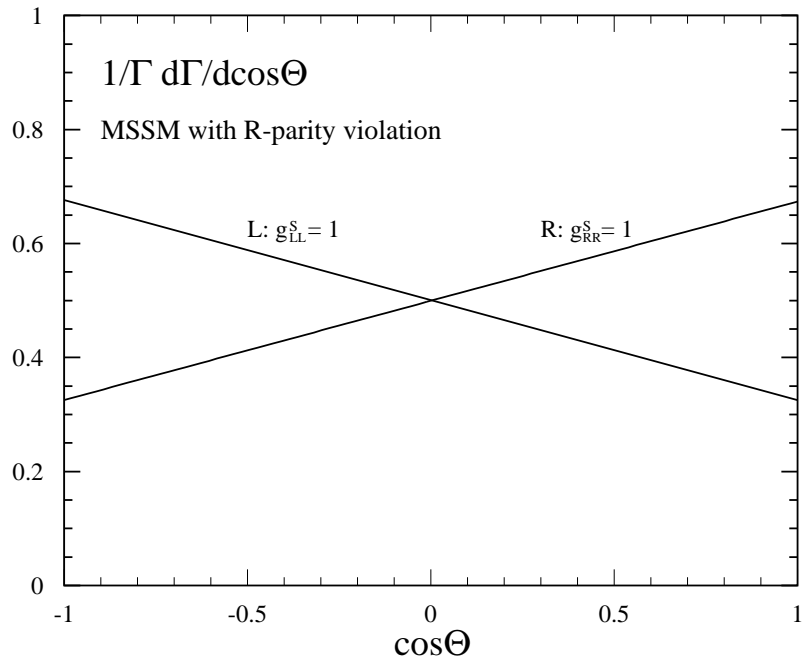}
\caption{$1/\Gamma$ $\differentiald\Gamma/\differentiald\cos\Theta$
  for the supersymmetric models with see--saw mechanism (upper figure)
  and $R$--parity violation (lower figure) discussed in
  Section~\ref{sec:models}. The results are obtained from a numerical
  calculation including all mass effects.}
\label{all_plots1}
\end{center}
\end{figure}

\begin{figure}
\vspace*{-10mm}
\begin{center}
\includegraphics[width=13cm]{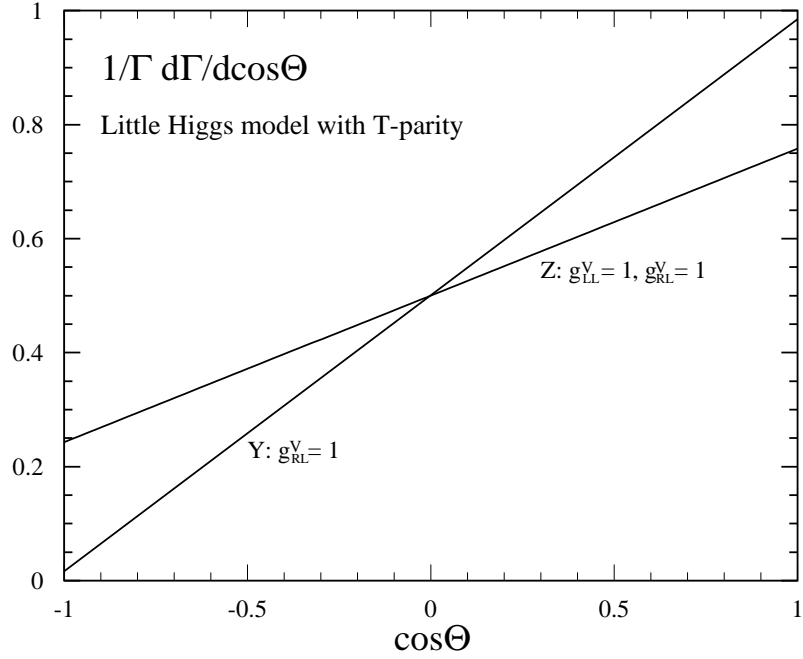}\\[-5mm]
\includegraphics[width=13cm]{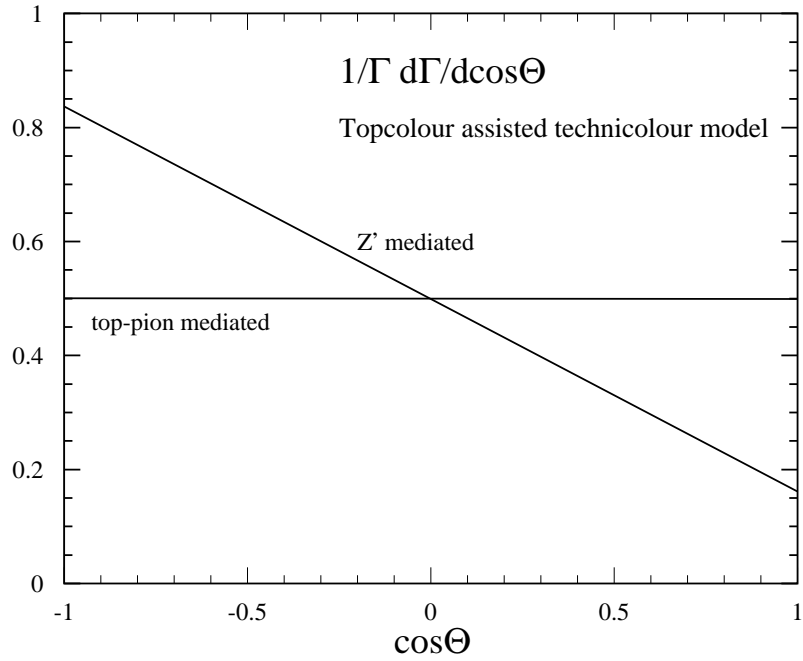}
\caption{$1/\Gamma$ $\differentiald\Gamma/\differentiald\cos\Theta$
  for the Littlest Higgs model (upper figure) and the
  topcolour--assisted technicolour model (lower figure) discussed in
  Section~\ref{sec:models}. The results are obtained from a numerical
  calculation including all mass effects.}
\label{all_plots2}
\end{center}
\end{figure}

\begin{figure}
\vspace*{-10mm}
\begin{center}
\includegraphics[width=13cm]{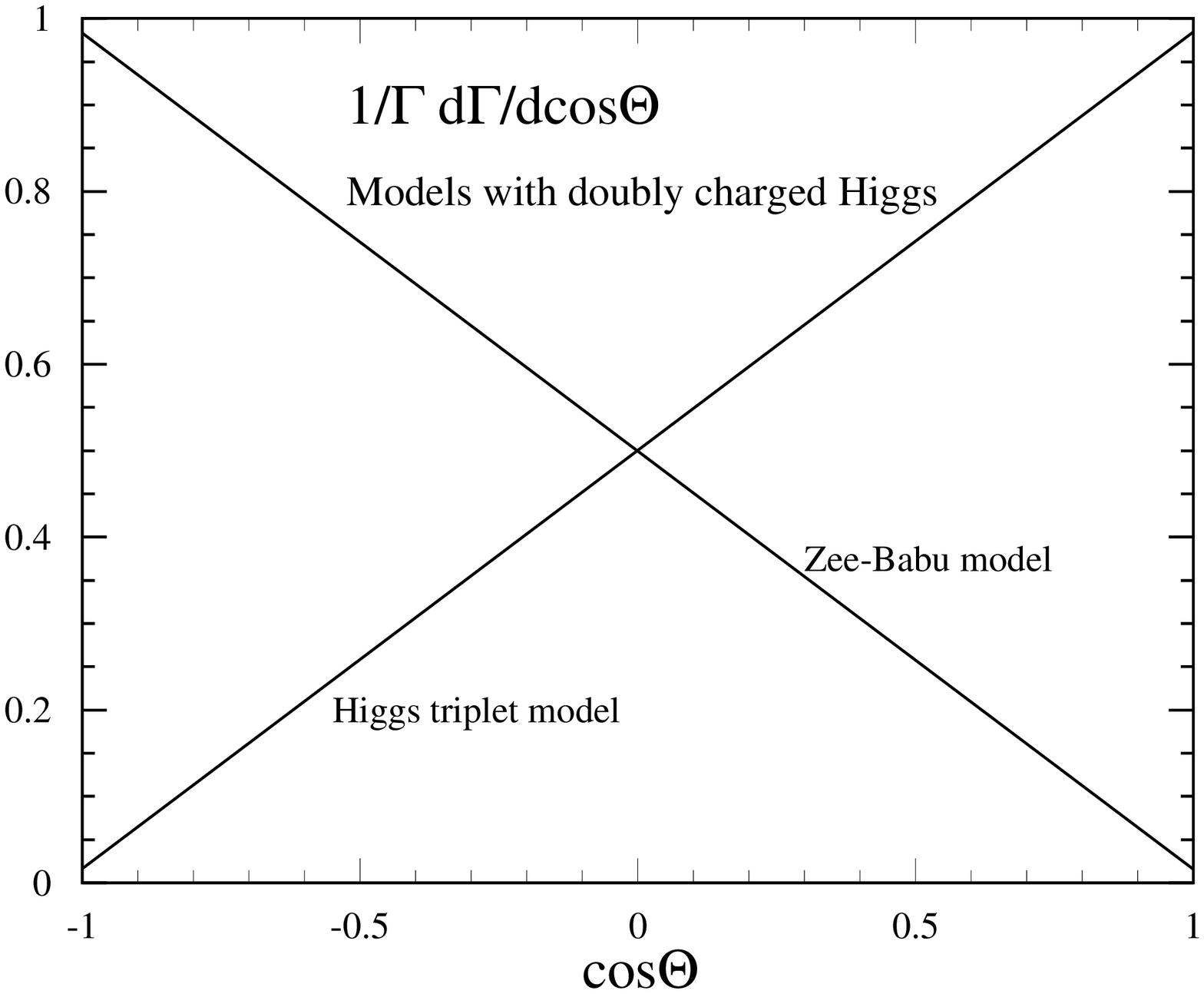}
\caption{$1/\Gamma$ $\differentiald\Gamma/\differentiald\cos\Theta$
  for the models with doubly--charged Higgs bosons discussed in
  Section~\ref{sec:models}. The results are obtained from a numerical
  calculation including all mass effects.}
\label{all_plots3}
\end{center}
\end{figure}

We summarize our results for the differential decay distributions $1
/ {\Gamma}$ ${\differentiald}{\Gamma} / {\differentiald} {\cos\Theta}$
within the various new physics models of Section~\ref{sec:models} in
Fig.~\ref{all_plots1}, \ref{all_plots2} and \ref{all_plots3} and
\citetable{slopes_table}.  [While Fig.~\ref{all_plots1},
\ref{all_plots2} and \ref{all_plots3} display the numerical results
including all mass effects, \citetable{slopes_table}\ quotes the
analytic results for massless muons, which are an excellent
approximation to the numerical calculation with full mass dependence.]
We can see that determining the slope to within $10\%$ should be
sufficient to distinguish between the see--saw MSSM, the Littlest
Higgs model with $T$--parity and the topcolour--assisted technicolour
model.  The $R$--parity--violating MSSM is not distinguishable from
the see--saw MSSM for ${\lambda}_{i22} {\lambda}_{i23}^{{\ast}}$ much
larger than ${\lambda}_{i32} {\lambda}_{i22}^{{\ast}}$, but is easily
discriminated from the Littlest Higgs model with $T$--parity and the
topcolour--assisted technicolour model, since the magnitude of the
slope is bounded by $1/6$, which is less than the slopes of the other
models.

With an estimated $2 {\times} 10^{8}$ \tauleptons\ produced from $W$
bosons in the first year of low--luminosity operation of the LHC, the
current upper bound on \tautothreemu\ means that even if the branching
ratio is just below the bound and the full detector efficiency is not
much lower than the simulated acceptance, we only expect about ten
events in the first twelve months.  However, after a year of
high--luminosity operation, we expect as much as a hundred events,
which should be sufficient to measure the \cosTheta\ dependence of the
decay width to within the desired $10\%$.

\begin{table} 
\begin{center}
\begin{tabular}{|l|c|}
\hline
\rule[0mm]{0mm}{0mm} Model   &\rule[-3mm]{0mm}{8mm} $\frac{\displaystyle
  1}{\displaystyle\Gamma}\frac{\displaystyle
  \differentiald\Gamma}{\displaystyle\differentiald
\cos\Theta} \propto 1 + A\cos\Theta$ 
\newhline
\rule[-3mm]{0mm}{8mm} MSSM with see--saw mechanism  & $A = 1/6$ 
\newhline
\rule[-3mm]{0mm}{8mm} MSSM with $R$--parity violation: &                          \\[1mm]
\rule[0mm]{0mm}{0mm} ``R'' (${\lambda}_{i22} {\lambda}_{i23}^{{\ast}} \gg {\lambda}_{i32}
{\lambda}_{i22}^{{\ast}}$) & $A = 1/6$   
\\[1mm] 
\rule[0mm]{0mm}{0mm} ``L'' (${\lambda}_{i22} {\lambda}_{i23}^{{\ast}} \ll {\lambda}_{i32}
{\lambda}_{i22}^{{\ast}}$) & $A = -1/6$ 
\newhline
\rule[-3mm]{0mm}{8mm} Littlest Higgs model with $T$--parity:         &    \\[1mm] 
\rule[0mm]{0mm}{0mm} ``Z'' (${\bar{Z}}_{{\text{odd}}}^{{\prime}{\tau}{\mu}} \gg {\bar{Y}}_{{\mu},{\text{odd}}}^{{\prime}{\tau}{\mu}}$) & $A = 5/18$
\\[1mm] 
\rule[0mm]{0mm}{0mm} ``Y'' (${\bar{Z}}_{{\text{odd}}}^{{\prime}{\tau}{\mu}} \ll
{\bar{Y}}_{{\mu},{\text{odd}}}^{{\prime}{\tau}{\mu}}$) & $A = 1/2$ 
\newhline
\rule[-3mm]{0mm}{8mm} Topcolour--assisted technicolour: &                      \\[1mm]
\rule[0mm]{0mm}{0mm} ``Z'' ($m_{{{\pi}_{t}}} {\gg} m_{{Z'}}$) & $A = -5/14$\\[1mm] 
\rule[0mm]{0mm}{0mm} ``P'' ($m_{{{\pi}_{t}}} {\ll} m_{{Z'}}$) & $A = 0$ 
\newhline
\rule[-3mm]{0mm}{8mm} Models with doubly--charged Higgs bosons: &  \\[1mm]
\rule[0mm]{0mm}{0mm} Higgs triplet model & $A = 1/2$\\[1mm]
\rule[0mm]{0mm}{0mm} Zee--Babu model  & $A = -1/2$ 
\newhline
\end{tabular}
\end{center}
\caption{The slope of 
  $1/\Gamma$
  $\differentiald\Gamma/\differentiald\cos\Theta$
  for the various new physics models discussed in
  Section~\ref{sec:models}. The numbers are obtained from the analytic
  results for massless muons.}
\label{slopes_table}
\end{table}

\section{Conclusion}

We have analyzed the lepton--flavour violating decay \tautothreemu\ 
and presented analytic expressions for the differential decay width
derived from an effective Lagrangian with general four--fermion
interactions. The results have been used in a Monte Carlo study to
estimate the experimental acceptance of \tautothreemu\ decays at the
LHC for generic sets of models.  We have derived specific predictions
for five classes of new physics models which predict the decay
\tautothreemu\ at rates that may be observable at the LHC: the
$R$--parity--conserving MSSM including see--saw neutrino masses at
large \texttanbeta, the $R$--parity--violating MSSM, the Littlest
Higgs model with $T$--parity, the topcolour--assisted technicolour
model, and models with doubly--charged Higgs bosons. For these models,
our Monte Carlo studies of the decay \tautothreemu\ at the LHC predict
experimental acceptances of approximately $25$--$30\%$. We have
emphasized that the models can be distinguished from each other by
measuring the angle between the $\tau$ polarization vector and the
momentum of the anti--muon.  This can be achieved within a year of
full luminosity at the LHC, if the branching ratio for the decay
\tautothreemu\ is close to its current upper bound.

\vspace*{5mm}

\noindent
{\bf Acknowledgments}\\
We would like to thank Tord Riemann for discussions on LFV violating
decays in the SM. The work of M.K.~and B.O'L.\ was supported in part
by the DFG grant KR~3345/1-1 and by the DFG SFB/TR9 ``Computational
Particle Physics''. The work of M.G. was supported in part by the BMBF
grant FSP2-CMS.

\end{document}